\documentclass[a4paper, aps, prl, longbibliography, twocolumn,nofootinbib]{revtex4-1}

\usepackage{asymptote} 
\usepackage{amsmath,amssymb,amsfonts}
\usepackage[colorlinks]{hyperref}
\usepackage{graphicx}
\graphicspath{{./pic/}}

\newcommand{\bw}{\begin{widetext}}
\newcommand{\ew}{\end{widetext}}
\newcommand{\be}{\begin{equation}}
\newcommand{\en}{\end{equation}}
\newcommand{\bee}{\begin{equation}}
\newcommand{\ene}{\end{equation}}
\newcommand{\bea}{\begin{eqnarray}}
\newcommand{\ena}{\end{eqnarray}}
\newcommand{\bes}{\begin{subequations}}
\newcommand{\ens}{\end{subequations}}
\newcommand{\bef}{\begin{figure}}
\newcommand{\enf}{\end{figure}}

\def\ie{{\it i.e.}}

\def\cala{\mathcal{A}}

\def\calc{\mathcal{C}}
\def\cald{\mathcal{D}}

\def\calh{\mathcal{H}}

\def\calk{\mathcal{K}}
\def\call{\mathcal{L}}
\def\calm{\mathcal{M}}

\def\calo{\mathcal{O}}
\def\calp{\mathcal{P}}

\def\calx{\mathcal{X}}

\def\to{\rightarrow}

\def\gev{{\rm GeV}}

\begin{document}


\title{Asymmetry Observables for Measuring Spin Correlations in Top-Quark Pair Production
}

\author{Kai Ma}
\email[Electronic address: ]{makainca@yeah.net}
\affiliation{School of Physics Science, Shaanxi University of Technology, Hanzhong 723000, Shaanxi, China}

\date{\today}

\begin{abstract}
Spin correlations between top-quark and anti-top-quark in pair production are sensitive to physics beyond the standard model. A complete set of observables, that can be used to measure all independent parameters of the production spin density matrix, is necessary to search for anomalous interactions in the production dynamics. In this letter, we provide a set of asymmetry observables for measuring all the density matrix elements. These observables are defined regardless of the production mechanism, and can be specified in any desired reference frame, particularly the laboratory frame in which the first evidence of spin correlation was observed experimentally. Our method can be used to find an optimal Lorentz frame  
in which the sensitivity to the density matrices is high for given production process and experimental environment, or to test possible correlations among the matrix elements.
\end{abstract}


\maketitle


%
Even through the observed scalar particle having mass $\sim125\gev$~\cite{ATLAS:CMS:2015} is likely the standard model (SM) Higgs boson, understanding the origin of electroweak symmetry breaking is still the main task of fundamental physics. Many extensions of the SM predict that the heavier the particle the stronger the interactions with unknown sector, which may shed light to the hierarchy problem~\cite{Hooft:1980} or the vacuum stability~\cite{Sher:1989, Degrassi:2012}. The top-quark being the heaviest particle that have been observed, is hence a natural probe of new physics above the electroweak scale~\cite{Fedor:2017,Markus:2017,Ulrich:2017}. Observable deviation of the top-quark property from the SM is, however, often constrained to be small~\cite{Markus:2017,Ulrich:2017}. Therefore, model-independent observables with high observability and discriminating power are desired.

The production rate of spin-polarized top-quark, which is predicted to be essentially zero in SM~\cite{Ulrich:2017}, is one of the most sensitive probes to deviation from the SM, and can be measured by studying the polar angle~\cite{Jezabek:1989} as well as the azimuthal angle distributions~\cite{Godbole:2010kr} of the decay products, particularly the lepton. On the other hand, the production spin correlations between the top-quark and anti-top-quark provide much more richer physics for probing anomalous interactions in the production dynamics~\cite{Aguilar-Saavedra:2014kpa}. The azimuthal angle difference between the two charged leptons in the laboratory frame was predicted in Ref.~\cite{Mahlon:2010gw} to be a sensitive observable, and has been observed in first in Ref.~\cite{ATLAS:2012ao}, and soon after in Refs.~\cite{Chatrchyan:2013wua, Aad:2014pwa}. In Refs.~\cite{Chatrchyan:2013wua, Aad:2014pwa, Aad:2013ksa, Aad:2014mfk, Aad:2015bfa}, the helicity correlation proposed in Ref.~\cite{Mahlon:1996} was also observed by measuring the correlation between the polar angle distributions of the leptons. In addition, the observable ``$S$ ratio" predicted in Ref.~\cite{Mahlon:2010gw} was also studied in Ref.~\cite{Aad:2014pwa}. There are also other related observables, for instance, the charge asymmetry proposed in Refs~\cite{Kuhn:1998,Kuhn:1999} as well as the charge lepton asymmetry in Refs.~\cite{Werner:2012,Berger:2012:A,Falkowski:2013}.

However, in order to get enough number of observables to extract possible new physics in the top-quark pair production dynamics, a complete set of observables, that can be used to measure all the production spin density matrix elements, is necessary. By choosing different quantization axes of the top-quarks, the authors in Ref.~\cite{Werne:2014} obtained 15 independent observables for a complete measurement. Even through these observables have been studied experimentally in Ref.~\cite{Aaboud:2016bit}, the requirement of momenta reconstruction for top-quarks can heavily dilute their sensitivity. For instance, it is the laboratory frame in which the first conclusive evidence of the production spin correlation was observed~\cite{Mahlon:2010gw, ATLAS:2012ao}, however, the corresponding observable can not be predicted directly by using the approach in Ref.~\cite{Werne:2014}. On the other hand, at the $e^{-}e^{+}$ collider, the reconstruction efficiency is much better in hadronic decay channels~\cite{Devetak:2011}, but there is not too much improvement in lepton-jet channels~ \cite{Rontsch:2015} and lepton-lepton channels~\cite{Aguilar:2015}. 

In view of above situation, it is promising to find a new complete set of observables, which can be directly accessible in the laboratory frame or any other desired reference frames that can be defined properly. In this letter, we provide a novel and systematic solution to this problem.
In contrast to the method in Ref.~\cite{Werne:2014}, the spins of both the top-quark pair and the lepton pair systems are projected along a common direction that can be defined in any desired reference frame. Spin correlations along other directions in the same or different reference frames can be easily obtained by direct Lorentz transformations.

For a general top-quark pair production process, $i\to X t \bar{t}$, where $i$ stands for the initial state of the colliding particles and $X$ is a set of other particles in addition to the top-quark pair, the production spin correlations can be affected in various ways. In this letter we confine ourselves to following extended interactions,
\bee\label{eq:LagGen}
\call = \call_{{\rm SM}} - \sum_{\alpha, i}\;\overline{\psi_{t}} \, \Gamma_{\alpha} \calo^{\alpha}(Y_{i}) \, \psi_{t} + {\rm h.c.}\,,
\ene
where $Y_{i}$ are possible new particles beyond the SM content, and $\Gamma_{\alpha}$ describes possible top-quark currents coupling to $Y_{i}$ through contractions with $\calo^{\alpha}(Y_{i})$ which in general can involve derivatives of the fields $Y_{i}$. $Y_{i}$ can also be particles of the set $X$. Neglecting higher order effects of the interactions in \eqref{eq:LagGen}, the decay properties of the top-quark are completely described by the SM. While our method can be applied equivalently for all decay channels (in SM), we consider only dileptonic decay channels, \ie, $t\to b\ell^{+}\nu_{\ell}$ and $\bar{t}\to\bar{b}\ell^{-}\bar{\nu}_{\ell}$, other channels will be briefly discussed at the end of this letter. 

No matter what is the dynamics in the top-quark pair production, by employing the narrow width approximation which while keeping the production spin correlations is an excellent simplification method, the transition amplitude of the above process can be written as a product of the production term and the decay term as follows,
\bee
\calm 
=
\sum_{ s_{1}, s_{2} = \pm1/2 } 
\calm_{P}(s_{1}, s_{2}) \calm_{D}(s_{1}, s_{2})\,,
\ene
where the narrow width approximation factors have been suppressed for convenience, and $s_{1}=\pm1/2$, $s_{2}=\pm1/2$ are helicity of top-quark and anti-top-quark, respectively. Correspondingly, the amplitude squared can also be factorized into production and decay density matrices,
\bee
\overline{ |\calm|^{2} } 
= \sum_{ s_{1}, s_{2}; s'_{1}, s'_{2} }  
\calp^{ s_{1}, s_{2} }_{ s'_{1}, s'_{2} }\;
\cald^{ s_{1}, s_{2} }_{ s'_{1}, s'_{2} }\,.
\ene
Spin correlations are encoded in the production spin density matrix $\calp^{ s_{1}, s_{2} }_{ s'_{1}, s'_{2} }$, and can be measured by studying angular distributions of the leptons which are determined by the decay density matrix $\cald^{ s_{1}, s_{2} }_{ s'_{1}, s'_{2} }$. However, apart from the machine-dependence of $\calp^{ s_{1}, s_{2} }_{ s'_{1}, s'_{2} }$, both the production and decay spin density matrices are also functions of the momenta of the top-quark pair, unless $\cald^{ s_{1}, s_{2} }_{ s'_{1}, s'_{2} }$ are calculated in the rest frames of top-quark and anti-top-quark simultaneously. Therefore, momenta reconstruction is unavoidable in this formalism. Furthermore, this mixture property also makes the optimization for getting as large as possible $\cald^{ s_{1}, s_{2} }_{ s'_{1}, s'_{2} }$ hard.

In this letter, we provide a novel approach to absorb the momenta dependence of the decay density matrix into the production density matrix by using spin projection along only one direction $\vec{Q}$ defined in a  common reference frame $R$ for all particles. In a chosen reference frame $R$, the top-quark pair and the lepton pair can be treated as two single systems whose spin can be either $0$ or $1$. The helicity of these two systems along the direction $\vec{Q}$ are equal, and can have values $\lambda_{f}={\rm s}, 0, \pm1$, where $\lambda_{f}={\rm s}$ means that the total spin is zero. After the spin projection, the amplitude squared $\overline{ |\calm|^{2} }$ can be decomposed as follows
\bee\label{eq:density}
\overline{ |\calm|^{2} } = \sum_{ \lambda_{f}, \lambda'_{f}= {\rm s}, 0, \pm1 }
\widetilde{\calp}^{ \lambda_{f} }_{ \lambda'_{f} } \;
\widetilde{\cald}^{ \lambda_{f} }_{ \lambda'_{f} }\,,
\ene
where $\widetilde{\cald}^{ \lambda_{f} }_{ \lambda'_{f}}$ are functions of only the angular variables of the leptons measured in a chosen reference frame $R$. The advantage of this factorization is that, as long as the relative direction of $\vec{Q}$ in $R$ is fixed, angular correlations determined by $\widetilde{\cald}^{ \lambda_{f} }_{ \lambda'_{f}}$ are universal in any reference frame. Hence, we can scan $\widetilde{\calp}^{ \lambda_{f} }_{ \lambda'_{f} }$ in all proper reference frames to obtain the most sensitive one. Alternatively, we can also change the projection direction $\vec{Q}$. This letter, in addition to provide a complete set of observables, the novel idea also facilitates the optimization. Our approach to the spin projection is explained as bellow.

The decay helicity amplitude $\calm_{D}(s_{1}, s_{2})$ can be written at leading order as a product of helicity amplitudes $\calm_{D_{1}}(s_{1})$ and $\calm_{D_{2}}(s_{s})$ for top-quark and anti-top-quark decays, respectively, which can be obtained directly according to the Feynman rules of the SM,
\bea
\calm_{D_{1}}(s_{1}) 
&=& 
\dfrac{ g_{W}^{2} }{2 D_{W_{1}} }  
\big\{\overline{u_{\nu_{\ell}}} \gamma_{\mu} P_{L} v_{ \bar{\ell} }\big\}
\big\{\overline{u_{b}}\gamma^{\mu} P_{L}u_{t}(s_{1})\big\}\,,
\\[2mm]
\calm_{D_{2}}(s_{2}) 
&=&
\dfrac{ g_{W}^{2} }{2 D_{W_{2}} }  
\big\{\overline{u_{ \ell }} \gamma_{\nu} P_{L} v_{\bar{\nu}_{\ell}}\big\}
\big\{\overline{v_{ \bar{t} }}(s_{2})\gamma^{\nu}P_{L}v_{ \bar{b} }\big\}\,,
\ena
where $D_{W_{1}}$ and $D_{W_{2}}$ are denominators of the propagators of $W$ bosons. 
Spin correlations between the anti-lepton (lepton) and the top-quark (anti-top-quark) can be clearly observed by applying Fierz transformations after replacing wave functions of the anti-lepton (lepton) and the neutrino (anti-neutrino) by the ones of their anti-particles \footnote{Note that the charge conjugation for leptons give $\overline{u_{\bar{\ell}}}$ and $v_{\nu_\ell}$ spinors for the final state $\bar{\ell}$ and $\nu_\ell$, respectively, and similarly for the $\ell + \bar{\nu}_\ell$ final state for $\bar{t}$ decays.},
\bea
\label{eq:TopDecay}
\calm_{D}(s_{1}) 
&=& 
-g_{W}^{2} D_{W_{1}}^{-1}
\big\{\overline{u_{\bar{\ell}}}\, P_{L}u_{t}(s_{1})\big\}
\big\{\overline{u_{b}}P_{R} v_{\nu_{\ell}}\big\}\,,
\\[2mm]
\label{eq:AopDecay}
\calm_{D}(s_{2}) 
&=& 
-g_{W}^{2} D_{W_{2}}^{-1}
\big\{\overline{v_{ \bar{t} }}(s_{2})P_{R} v_{\ell}\big\}
\big\{\overline{u_{\bar{\nu}_{\ell}} } P_{L} v_{\bar{b}} \big\}\,.
\ena
Factorization of the lepton pair system from top-quark pair system can be realized by simply applying the Fierz transformation one more step. Then we find, 
\bee
\label{eq:ToponiaDecay}
\calm(s_{1}, s_{2}) 
= \calh^{\mu}(s_{1}, s_{2}) \,\calk_{\mu} \; \calx \,\overline{\calx} \,,
\ene
where
\bea
\label{eq:decay:HEL:Fact}
\calh^{\mu}(s_{1}, s_{2}) 
&=&
\overline{v_{\bar{t}}}(s_{2})\gamma^{\mu}P_{L}u_{t}(s_{1})\,,
\\[2mm]
\calk_{\mu}
&=&
-\dfrac{1}{2} \overline{u_{\ell}}\,\gamma_{\mu}P_{L} v_{\bar{\ell}} \,,
\\[2mm]
\calx
&=&
g_{W}^{2} D_{W_{2}}^{-1}
\big[\overline{u_{b}}P_{R} v_{\nu_{\ell}}\big] \,,
\\[2mm]
\overline{\calx} 
&=&
g_{W}^{2} D_{W_{1}}^{-1}
\big[\overline{u_{\bar{\nu}_{\ell} }} P_{L} v_{\bar{b}} \big]\,.
\ena
One can observe that, apart from the factors $\calx$ and $\overline{\calx}$, the decay helicity amplitude is written as a Lorentz contraction between the currents generated by the top-quark pair and lepton pair. Spin projection along chosen direction $\vec{Q}$ in a reference frame $R$ can be easily obtained by inserting a complete projection relation, 
\bee
g_{\mu\nu} 
= \sum_{\lambda_{f}={\rm s}, 0, \pm} 
\eta_{\lambda_{f}}\epsilon_{\mu}^{\ast}(\vec{Q}, \lambda_{f})
\epsilon_{\nu}(\vec{Q}, \lambda_{f})\,,
\ene 
where $\eta_{{\rm s}}=1$ and $\eta_{\lambda_{f}=0,\pm1}=-1$, into the helicity amplitude \eqref{eq:ToponiaDecay}. Then we find,
\bee
\calm(s_{1}, s_{2})  
= 
\sum_{\lambda_{f}=s, 0, \pm}
\widetilde{\calh}_{\lambda_{f}}(s_{1}, s_{2}) \,\widetilde{\calk}_{\lambda_{f}} \,,
\ene
where the scalar functions $\widetilde{\calh}_{\lambda_{f}}(s_{1}, s_{2})$ and 
$\widetilde{\calk}_{\lambda_{f}}$ are defined as 
\bea
\widetilde{\calh}_{\lambda_{f}}(s_{1}, s_{2})
&=& 
\sqrt{ E_{1}E_{2} }\, \calx \,\overline{\calx} \, \calh^{\mu}(s_{1}, s_{2}) \, \epsilon_{\mu}^{\ast}(\vec{Q}, \lambda_{f})\,, 
\\[2mm]
\widetilde{\calk}_{\lambda_{f}} 
&=&
\frac{1}{\sqrt{ E_{1}E_{2} } }\; \eta_{\lambda_{f}}\calk^{\nu} \epsilon_{\nu}(\vec{Q}, \lambda_{f})\,,
\ena
with $E_{1}$ and $E_{2}$ are energy of the anti-lepton and lepton in $R$, respectively. The spin-projected decay density matrix is simply given as
\bee
\widetilde{\cald}^{ \lambda_{f} }_{ \lambda'_{f} } 
=
\widetilde{\calk}_{\lambda_{f}} \widetilde{\calk}^{\dag}_{\lambda'_{f}}\,.
\ene 
And the spin-projected production density matrix can be obtained by summing up the helicity $s_{1}$ and $s_{2}$, 
\bee
\widetilde{\calp}^{ \lambda_{f} }_{ \lambda'_{f} } 
=
\sum_{ s_{1}, s_{2};\, s'_{1}, s'_{2} }
\calp^{ s_{1}, s_{2} }_{ s'_{1}, s'_{2} }\;
\widetilde{\calh}_{\lambda_{f}}(s_{1}, s_{2}) \;
\widetilde{\calh}^{\dag}_{\lambda'_{f}}(s'_{1}, s'_{2})\,.
\ene
Clearly, nontrivial angular distributions of the lepton and anti-lepton are completely encoded in the projected density matrix $\widetilde{\cald}^{ \lambda_{f} }_{ \lambda'_{f} }$. Spin correlations of the top-quark pair coming from either the SM dynamics or the anomalous interaction in \eqref{eq:LagGen} are described by the projected density matrix $\widetilde{\calp}^{ \lambda_{f} }_{ \lambda'_{f} }$, and then affect the angular distributions the lepton and anti-lepton, and therefore can be measured by observing the angular distributions.

Explicit expressions of $\widetilde{\cald}^{ \lambda_{f} }_{ \lambda'_{f} }$ can be obtained in a straightforward way once $R$ and $\vec{Q}$ are given. Without loss of generality, we can set $\vec{Q}= (0, 0, 1)$, \ie, it is the unite vector along the $z$ direction in $R$, then we find 
\bee\label{eq:KinDecay}
\widetilde{\calk}_{\lambda_{f}}
=
\begin{cases}
-\sqrt{2}\, c_{1}s_{2} e^{i\phi_{+} } & \lambda_{f} = +1
\\[3.8mm]
\sqrt{2} \, s_{1}c_{2} e^{- i\phi_{+}  }  & \lambda_{f} = -1
\\[3.8mm]
c_{1}c_{2} e^{i \phi_{-} } - s_{1}s_{2} e^{-i \phi_{-} } & \lambda_{f} = 0
\\[3.8mm]
c_{1}c_{2} e^{i \phi_{-} } + s_{1}s_{2} e^{-i \phi_{-} } & \lambda_{f} = {\rm s}
\end{cases}
\,,
\ene
where $\theta_{1}$ ($\phi_{1}$) and $\theta_{2}$ ($\phi_{2}$) are the polar (azimuthal) angles of the anti-lepton and lepton in the reference frame $R$, respectively; 
and $c_{i} = \cos(\theta_{i}/2)$, $s_{i} = \sin(\theta_{i}/2)$, and the
phases $\phi_{\pm} = (\phi_{1}\pm\phi_{2})/2$. 

Spin correlations between the lepton and top-quark systems are encoded in the following density matrix,
\bee
\rho^{\lambda_{f}}_{\lambda'_{f}}
=
\widetilde{\calp}^{ \lambda_{f} }_{ \lambda'_{f} }
\widetilde{\cald}^{ \lambda_{f} }_{ \lambda'_{f} }\,,
\ene
whose matrix elements are given as,
\bea
\rho^{+}_{+} 
&=& 
\dfrac{1}{2} \widetilde{\calp}^{+}_{+} 
(1 + \cos\theta_{1})(1 - \cos\theta_{2})\,,
\\[2mm]
\rho^{-}_{-} 
&=& 
\dfrac{1}{2} \widetilde{\calp}^{-}_{-}  
(1 - \cos\theta_{1})(1 + \cos\theta_{2})\,,
\\[2mm]
\rho^{0}_{0} 
&=& 
\dfrac{1}{2} \widetilde{\calp}^{0}_{0} 
\big( 1 + \cos\theta_{1}\cos\theta_{2} -
\sin\theta_{1}\sin\theta_{2}\cos(2\phi_{-})\big)\,,
\\[2mm]
\rho^{s}_{s} 
&=& 
\dfrac{1}{2} \widetilde{\calp}^{{\rm s}}_{{\rm s}} 
\big( 1 + \cos\theta_{1}\cos\theta_{2} +
\sin\theta_{1}\sin\theta_{2}\cos(2\phi_{-})\big)\,,
\ena
\begin{widetext}
\bea
\label{eq:Int:PM} {\rm Re}\big[\rho^{+}_{-}\big]
&=& 
- \dfrac{1}{2} \big|\widetilde{\calp}^{+}_{-} \big|
\sin\theta_{1}\sin\theta_{2}\cos(2\phi_{+} + \delta^{+}_{-})\,,
\\[2mm]
\label{eq:Int:PZ} {\rm Re}\big[\rho^{+}_{0}\big]
&=& 
\dfrac{1}{2\sqrt{2}} \big|\widetilde{\calp}^{+}_{0}\big|  
\bigg\{ 
(1-\cos\theta_{2})\sin\theta_{1} \cos(\phi_{1} + \delta^{+}_{0} )
- (1+\cos\theta_{1})\sin\theta_{2} \cos(\phi_{2}+ \delta^{+}_{0})
\bigg\}\,,
\\[2mm]
\label{eq:Int:PS} {\rm Re}\big[\rho^{+}_{s}\big]
&=& 
\dfrac{-1}{2\sqrt{2}} \big|\widetilde{\calp}^{+}_{{\rm s}} \big| 
\bigg\{
(1-\cos\theta_{2})\sin\theta_{1} \cos(\phi_{1} + \delta^{+}_{{\rm s}} )
+ (1+\cos\theta_{1})\sin\theta_{2} \cos(\phi_{2}+ \delta^{+}_{{\rm s}})
\bigg\}\,,
\\[2mm]
\label{eq:Int:MZ} {\rm Re}\big[\rho^{-}_{0}\big]
&=& 
\dfrac{1}{2\sqrt{2}} \big|\widetilde{\calp}^{-}_{0}\big| 
\bigg\{
(1 + \cos\theta_{2})\sin\theta_{1} \cos(\phi_{1} - \delta^{-}_{0} )
- (1- \cos\theta_{1})\sin\theta_{2} \cos(\phi_{2} - \delta^{-}_{0})
\bigg\}\,,
\\[2mm]
\label{eq:Int:MS} {\rm Re}\big[\rho^{-}_{{\rm s}}\big]
&=& 
\dfrac{1}{2\sqrt{2}} \big|\widetilde{\calp}^{-}_{{\rm s}}\big|  
\bigg\{ 
(1 + \cos\theta_{2})\sin\theta_{1} \cos(\phi_{1} - \delta^{-}_{{\rm s}} )
+ (1- \cos\theta_{1})\sin\theta_{2} \cos(\phi_{2} - \delta^{-}_{{\rm s}})
\bigg\}\,,
\\[2mm]
\label{eq:Int:ZS} {\rm Re}\big[\rho^{0}_{{\rm s}}\big]
&=& 
\dfrac{1}{ 2 } \big|\widetilde{\calp}^{0}_{{\rm s}} \big|
\bigg\{ 
(\cos\theta_{1} + \cos\theta_{2} ) \cos(\delta^{0}_{{\rm s}} )
- \sin\theta_{1}\sin\theta_{2} \sin(2\phi_{-}) \sin(\delta^{0}_{{\rm s}} )
\bigg\}\,,
\ena
\end{widetext}
where $\delta^{\lambda_{f}}_{\lambda'_{f}}$ are phases of the projected density matrix $\widetilde{\calp}^{ \lambda_{f} }_{ \lambda'_{f} }$, and in general depend on both the production dynamics and kinematics.
From the above matrix elements, we can directly obtain the desired angular distributions, and hence  nontrivial angular correlations. For instances, the observable proposed in Ref.~\cite{Godbole:2010kr} is related to the elements $\rho^{\pm}_{0}$ and $\rho^{\pm}_{{\rm s}}$; the observable predicted in Ref.~\cite{Mahlon:2010gw} and measured in Refs.~\cite{ATLAS:2012ao,Chatrchyan:2013wua, Aad:2014pwa} is related to the diagonal elements $\rho^{0}_{0}$ and $\rho^{{\rm s}}_{{\rm s}}$, as well as the off-diagonal element $\rho^{0}_{{\rm s}}$; the helicity correlation in Ref.~\cite{Mahlon:1996} are given by the diagonal elements; the $CP$ violation observables proposed in Ref.~\cite{Ma:2016:Toponia} are related to the elements $\rho^{\pm}_{0}$ and $\rho^{\pm}_{{\rm s}}$, as well as the element $\rho^{+}_{-}$ (in the $t\bar{t}$ rest frame); 

Most interestingly, the 15 independent observables proposed in Ref.~\cite{Werne:2014} are related to the diagonal elements projected along three different directions in the $t\bar{t}$ rest frame. In contrast, our approach can provide 15 independent observables by using only one projection direction. First of all, the production cross section, which is spin-independent, is given by the diagonal elements,
\bee
\sigma 
= 
\frac{1}{2} \bigg( \overline{ \widetilde{\calp}^{{\rm s}}_{{\rm s}} } + \overline{\widetilde{\calp}^{0}_{0}} + \overline{\widetilde{\calp}^{+}_{+}} + \overline{\widetilde{\calp}^{-}_{-}} \bigg)\,,
\ene
where the over line ``\,$\overline{~^{~}}$\," means summing up PDFs of the initial state and  integrating over phase space of the final states except for the angular variables of the lepton pair which have been integrated out explicitly with following normalization convention,
\bee
d\Phi_{2} 
= 
\frac{1}{16\pi^{2}}\int_{-1}^{1} d\cos\theta_{1}  \int_{0}^{2\pi} d\phi_{1}
\int_{-1}^{1} d\cos\theta_{2}  \int_{0}^{2\pi} d\phi_{2}\,.
\ene 
For the spin-dependent observables, we define following two kinds of asymmetries. The general definition of the first kind of observables, that we call them even asymmetries, is given as 
\bee
\cala\big[ f( \varsigma) \big] 
= \frac{ \sigma\big[ g_{+}( \varsigma) > 0 \big] - \sigma\big[ g_{+}(\varsigma) < 0 \big] }{  \sigma\big[ g_{+}( \varsigma) > 0 \big] + \sigma\big[ g_{+}(\varsigma) < 0 \big]  }\,,
\ene
where $f(\varsigma)$ is a function of the angular variables $\varsigma=\{\theta_{i}, \phi_{i}\}$, and $g_{+}(\varsigma) = f( \varsigma) + f( -\varsigma)$. We call the second kind of observables as odd asymmetries, and the general definition is given as
\bee
\calc\big[ f( \varsigma) \big] 
= \frac{ \sigma\big[ g_{-}( \varsigma) > 0 \big] - \sigma\big[ g_{-}(\varsigma) < 0 \big] }{  \sigma\big[ g_{-}( \varsigma) > 0 \big] + \sigma\big[ g_{-}(\varsigma) < 0 \big]  }\,,
\ene
where $g_{-}( \varsigma) = f( \varsigma) - f( - \varsigma)$. The instances of these two kinds of asymmetries can be obtained by studying the matrix elements $\rho^{\lambda_{f}}_{\lambda'_{f}}$.

{\bf Even Asymmetries:} 
The simplest observables are $\cala_{\theta_{i}}= \cala[ \cos\theta_{i} ]$ for the single side distributions, which measure the polarizations of top-quarks inclusively,
\bee
\cala_{\theta_{i}}
=
\dfrac{1}{4\sigma}\, \bigg\{  
Q_{i} \big(\overline{ \calp^{-}_{-} } - \overline{ \calp^{+}_{+} } \big)  
+ 2\overline{ \calp^{0}_{{\rm s}}  \cos\delta^{0}_{{\rm s}}}
\bigg\} \,,
\ene 
where $Q_{i=1,2}$ are electric charges of the anti-lepton and lepton, respectively, in unite of $|e|$. There are also single side even asymmetries for the azimuthal angles, $\cala_{ \phi_{i} } = \cala[ \cos\phi_{i} ]$ having following explicit expressions,
\bee
\cala_{ \phi_{i} }=
\dfrac{ -1 }{2\sqrt{2}\sigma}\, \sum_{\lambda=\pm1}
\bigg\{ \lambda \overline{ \calp^{\lambda}_{{\rm s}}  \cos\delta^{\lambda}_{{\rm s}} }
 + Q_{i} \overline{ \calp^{\lambda}_{0}  \cos\delta^{\lambda}_{0} } \bigg\}\,.
\ene
$\cala_{\theta_{1}\theta_{2}}=\cala[ \cos\theta_{1}\cos\theta_{2} ]$ is the simplest 
double side even asymmetry, and measures helicity correlation,
\bee
\cala_{\theta_{1}\theta_{2}}
=
\dfrac{1}{8\sigma}\,
\bigg\{ \overline{ \calp^{{\rm s}}_{{\rm s}} } + \overline{ \calp^{0}_{0} } - \overline{ \calp^{+}_{+} } - \overline{ \calp^{-}_{-} } \bigg\}\,.
\ene
Furthermore, linear combinations of the azimuthal angles can also be used to define double side asymmetries as $\cala_{ \phi_{\pm} } = \cala[ \cos(2\phi_{\pm}) ]$. And the explicit expressions are,
\bea
\cala_{ \phi_{+} }
&=&
\dfrac{ \pi }{4\sigma}\, \overline{\calp^{+}_{-}  \cos \delta^{+}_{-} }\,,
\\[2mm]
\cala_{ \phi_{-} }
&=&
\dfrac{ \pi }{8\sigma}\,
\bigg\{ \overline{ \calp^{0}_{0} }  - \overline{ \calp^{{\rm s}}_{{\rm s}} } \bigg\}\,.
\ena
There are also double side even asymmetries involving polar and azimuthal angles of oppositely charged leptons, $\cala_{\theta_{i} \phi_{i'} } = \cala[ \cos(\theta_{i})\cos(\phi_{i'}) ] $ with $i\neq i'$,
\bee
\cala_{\theta_{i} \phi_{i'} }
=
\dfrac{ 1 }{4\sqrt{2}\sigma}\, \sum_{\lambda=\pm1}
\bigg\{ Q_{i} \overline{ \calp^{\lambda}_{{\rm s}}   \cos\delta^{\lambda}_{{\rm s} } } - 
\lambda \overline{ \calp^{\lambda}_{0}  \cos\delta^{\lambda}_{0} }  \bigg\}\,.
\ene
In total we have 9 even asymmetry observables, and it is 10 when the total cross section is included.
In case of that all the phases are (averagely) zero, the above observables are enough to determine the matrix elements. However, in general odd asymmetry observables are we necessary to determine the phases.

{\bf Odd Asymmetries:}  
Similarly, there are both single side and double side odd asymmetries. The simplest single side odd asymmetries are $\calc_{\phi_{i}} = \calc\big[ \sin\phi_{i} \big]$,
\bee
\calc_{\phi_{i}}
= \frac{ 1  }{ 2 \sqrt{2} \sigma } 
\sum_{\lambda=\pm1} \bigg\{ 
\overline{ \calp^{ \lambda }_{{\rm s}}  \sin\delta^{\lambda}_{{\rm s}} }
+
Q_{i} \lambda \, \overline{ \calp^{ \lambda }_{0}  \sin\delta^{\lambda}_{0}  }
\bigg\}\,.
\ene
In contrast, there are no single side odd asymmetry for polar angles. On the other hand,
linearly combinations of the azimuthal angles can give two double side asymmetries $\calc_{\phi_{+}} = \calc[ \sin(2\phi_{+}) ]$ and $\calc_{\phi_{-}} = \calc[ \sin(2\phi_{-}) ]$ as follows,
\bea
\calc_{\phi_{+}}
&=&
-\dfrac{ \pi }{8\sigma}\,\overline{ \calp^{+}_{-}  \, \sin \delta^{+}_{-} }   \,,
\\[2mm]
\calc_{\phi_{-}}
&=&\;\;\;
\dfrac{ \pi }{8\sigma}\,\overline{ \calp^{0}_{{\rm s}}  \, \sin\delta^{0}_{{\rm s}} }  \,.
\ena
In addition, there are two more double side odd asymmetries involving azimuthal and polar angles of oppositely charged particles, $\calc_{\theta_{i} \phi_{i'} } = \calc[ \cos(\theta_{i})\sin(\phi_{i'}) ] $ with $i\neq i'$, 
\bee
\calc_{\theta_{i} \phi_{i'} }
=
\dfrac{ 1 }{4\sqrt{2}\sigma}\, \sum_{\lambda=\pm1}
\bigg\{ \overline{ \calp^{\lambda}_{0}  \sin\delta^{\lambda}_{0} } 
- Q_{i} \lambda \overline{ \calp^{\lambda}_{{\rm s}}   \sin\delta^{\lambda}_{{\rm s} } } 
 \bigg\}\,.
\ene
In total, there are 6 odd asymmetry observables. Together with the 9 even asymmetries, 15 independent observables are obtained, and correspond to 15 independent parameters of a normalized Hermitian density matrix. 

However, the 15 matrix elements (16 before normalization) may not be independent in general. If the  top-quark pair are generated in pure quantum state, the production density matrix can be obtained by using transition amplitude, $\widetilde{\calp}^{ \lambda_{t} }_{ \lambda'_{t} }=\widetilde{\calm}_{P}( \lambda_{t}) \widetilde{\calm}_{P}^{\dag}( \lambda'_{t})$. Then there are only $4$ independent magnitudes $\big|\calm_{P}( \lambda_{t})\big|$, and following relations have to be hold exactly,
\bee\label{eq:rel:mag}
\left| \widetilde{\calp}^{ \lambda_{t} }_{ \lambda'_{t} } \right|
= \sqrt { \widetilde{\calp}^{ \lambda_{t} }_{ \lambda_{t} } \;  
\widetilde{\calp}^{ \lambda'_{t} }_{ \lambda'_{t} }  }\,.
\ene
Note that this number reduce to 3 after normalization. Furthermore, the number of independent phases is reduced to 3 (after an overall phase is removed), and following relations are also exact,
\bee\label{eq:rel:phase}
\delta^{\lambda_{t}}_{\lambda'_{t} } 
=  \delta( \lambda_{t} ) - \delta( \lambda'_{t} ) 
=\delta^{\lambda_{t}}_{\lambda''_{t} } - \delta^{\lambda''_{t}}_{\lambda'_{t} }\,,
\ene
where $\delta( \lambda_{t} )$ are phases of the production helicity amplitudes $\widetilde{\calm}_{P}( \lambda_{t})$. At hadron collider, for instance LHC, the colliding protons are mixed states, and the mixture are described by the PDFs. As a result the top-quark pair are not in pure state, and hence the relations \eqref{eq:rel:mag} and \eqref{eq:rel:phase} can be violated. However, if only one kind of the subprocess dominates the transition under considering, then the relations should be hold approximately. At LHC, the top-quark pair are produced dominantly through the $gg$ subprocess. Therefore, it is interesting to test the relations \eqref{eq:rel:mag} and \eqref{eq:rel:phase} by using spin correlation observables. Comparing to the observables proposed in Ref.~\cite{Werne:2014} that are defined in three different spin polarization directions, it is more easier to use our observables to test above relations.

In summary, we have provided a novel method to study the spin correlations between top-quark and anti-top-quark in top-quark pair production, and can be easily used to test the SM and probe possible new physics in the production dynamics. Our method is model-independent under the assumption that decays of top are completely described by the SM. Possible non-SM physics contribution in the decay part can be treated as a part of  uncertainties of our asymmetry observables in all the measurements on new physics effects in the production part. Because measurements on the top decay properties are improving rapidly, it is expected that the uncertainty due to potential non-SM top decays will become smaller in the future.
Based on a simple factorization of the decay helicity amplitude, the top-quark pair and the lepton pair can be treated as two single systems, and hence the production spin correlations can be studied easily by using angular distributions of the leptons. We have proposed 15 independent asymmetry observables to completely measure the production spin density matrix. All these observables are referring only one spin polarization direction in a chosen reference frame which can be the laboratory frame or any other desired reference frame that can be defined properly. Therefore, optimization of the observables for concrete models can be easily done. On the other hand, in order to precisely measure the physical property of top, the SM prediction on these observables have to be calculated accurately. We leave this part of studies in future works. Furthermore, in case of that the $W$s decay hadronically, our calculations are still valid, and hence similar asymmetries can also be defined in principle. Since measuring electric charge of the QCD jet is very hard, sensitivity of these observables can be strongly diluted, but is expected to be not too worse in case of that only one of the $W$s decay into jet.

\section*{Acknowledgements}
K.M. thanks very much Kaoru Hagiwara, Xinmin Zhang, Fa Peng Huang and She-Sheng Xue for useful discussions. This study is supported by the China Scholarship Council, and the National Natural Science Foundation of China under Grant No. 11647018.

\bibliography{Phenomenology}

\end{document}